\documentclass{llncs}

\usepackage{url}
\usepackage{multirow}
\usepackage{listings}
\usepackage{array}    
\usepackage{verbatim} 
\usepackage{wrapfig}
\usepackage[final]{microtype}
\usepackage{footnote}

\usepackage[usenames, dvipsnames]{color}
\usepackage[dvipsnames]{xcolor}
\usepackage{pifont}
\usepackage[fleqn]{amsmath}
\usepackage{amssymb}
\usepackage{tabularx}

\makesavenoteenv{tabular}

\lstnewenvironment{code}[1][]
  {\noindent
   \vspace{-0.5\baselineskip}
   \lstset{basicstyle=\ttfamily,
           frame=single,
           language=Haskell,
           keywordstyle=\color{black},
           #1}
   \fontsize{8pt}{8pt}\selectfont}
  {}

\newcommand{\cmark}{\ding{51}}
\newcommand{\xmark}{\ding{55}}
\newcommand{\pass}{\color{ForestGreen}\cmark}

\newcommand{\fail}{\color{red}\xmark}

\newcommand{\expectfail}{\color{orange}\textbf{!}}

\newcommand{\associativity}[1]{$(a #1 b) #1 c = a #1 (b #1 c)$}

\newcommand{\identitytwo}[2]{$#2 #1 a = a #1 #2 = a$}

\newcommand{\commutativity}[1]{$a #1 b = b #1 a$}
\newcommand{\distributivityleft}[2]{$a #1 (b #2 c) = (a #1 b) #2 (a #1 c)$}
\newcommand{\distributivityright}[2]{$(a #2 b) #1 c = (a #1 c) #2 (b #1 c)$}
\newcommand{\annihilation}[2]{$#2 #1 a = a #1 #2 = #2$}

\newcommand\NOTE[1]{} 

\begin{document}

\title{Proving Type Class Laws for Haskell}

\author{Andreas Arvidsson \and Moa Johansson \and Robin Touche}
\institute{Department of Computer Science and Engineering, Chalmers University of Technology
	\email{andreas.arvidson@gmail.com, moa.johansson@chalmers.se, robin\_touche@hotmail.com}\footnote{This paper describes work from Anders Arvidsson's and Robin Touche's joint MSc thesis at Chalmers \cite{msc}, supervised by Moa Johansson.}
	}


\maketitle

\begin{abstract}
Type classes in Haskell are used to implement ad-hoc polymorphism, i.e. a way to ensure both to the programmer and the compiler that a set of functions are defined for a specific data type. All instances of such type classes are expected to behave in a certain way and satisfy laws associated with the respective class. These are however typically just stated in comments and as such, there is no real way to enforce that they hold. In this paper we describe 
a system which allows the user to write down type class laws which are then automatically instantiated and sent to an inductive theorem prover when declaring a new instance of a type class.
\end{abstract}

\section{Introduction}
Type classes in Haskell are used to implement ad-hoc polymorphism, or overloading \cite{Wadler89}. 
They allow programmers to define functions which behave differently depending on the types of their arguments. A type-class declares a set of (abstract) functions and their types, and any datatype declared an instance of the class has to provide implementations for those functions. Below is the Haskell library type class \texttt{Monoid}, which declares the three abstract functions \texttt{mempty}, \texttt{mappend} and \texttt{mconcat}:

\begin{code}
class Monoid a where 
	mempty :: a
	mappend :: a -> a -> a
	mconcat :: [a] -> a
\end{code}

Usually, when we define a type-class we have some expectation on what properties instances of this type class should satisfy when implementing the functions specified. These are captured by \emph{type class laws}. For monoids, any instance is supposed to satisfy the laws below: 

\clearpage
\begin{code}
-- Left identity
mappend mempty x = x 

-- Right identity
mappend x mempty = x

-- Associativity of mappend
mappend x (mappend y z) = mappend (mappend x y) z

-- Specification of mconcat
mconcat = foldr mappend mempty
\end{code}
The last law is in fact a specification for the \emph{default implementation} of the \texttt{mconcat} function, which is commonly used (unless the user wants to provide their own, optimised, implementation). The last law then becomes a trivial identity.


\example
\label{ex:natmonoid}
The most obvious instance of \texttt{Monoid} is probably \texttt{Lists}, but we may for instance also declare the natural numbers to be monoids, with \texttt{+} corresponding to \texttt{mappend}:

\begin{code}
data Nat = Zero | Succ Nat

-- Natural number addition
(+) :: Nat -> Nat -> Nat
Zero + a = a
(Succ a) + b = Succ (a + b)

-- Natural number multiplication
(*) :: Nat -> Nat -> Nat
Zero * m = Zero
(Succ n) * m = m + (n * m)

-- Make Nat an instance of the Monoid type class
instance Monoid Nat where
 	mempty = Zero
	mappend = (+)
\end{code}
We could also have declared \texttt{Nat} a monoid in a different manner, with with \texttt{*} corresponding to \texttt{mappend}:

\begin{code}
instance Monoid Nat where
 	mempty = Succ Zero
	mappend = (*)
\end{code}
These instances of the \texttt{Monoid} class are quite simple. By just looking at them, we might convince ourselves that they behave in accordance with the type class laws for monoids, and settle for that. But what if we had a more complicated instance or had made a mistake? Unfortunately, in Haskell, type class laws are typically only stated in comments or documentation, if at all, and there is no support for checking that an instance of a type class actually behaves in accordance with the laws. Furthermore, type class laws could be used in, for example, compiler optimisations. Any violation of the laws could then cause inconsistencies between the original code and the optimised version, which is clearly undesirable.  

To address these problems, Jeuring et al. developed a framework for expressing and testing type class laws in Haskell using the  QuickCheck tool \cite{Jeuring2012,quickcheck}. As further work, they identify the need to also provide stronger guarantees by also \emph{proving} type class laws using an automated theorem prover. However, the type class laws typically present in Haskell programs often involve recursive functions and datatypes, which means that we might need induction to prove them. While there has been much success using SMT-solvers and first-order theorem provers for reasoning about programs, such provers, e.g. Z3 and E \cite{z3,eprover},  typically do not support induction. Some of the difficulties with inductive proofs is that they often require auxiliary lemmas, which themselves require induction. A system built to handle these kind of problems is HipSpec \cite{hipspecCADE}, a state-of-the-art automated inductive theorem for Haskell. 
Our contributions combine the ideas from earlier work on testing type class laws with inductive theorem proving, and allow us to:
\begin{itemize}
\item Write down type class laws in Haskell as \emph{abstract properties} (\S \ref{sec:expressing}), including support for types with class constraints. 

\item Automatically instantiate these abstract properties when new instances of a type class is declared, and translate them into a intermediate language called TIP \cite{tip}, suitable for passing on to automated theorem provers (\S \ref{sec:spec}).

\item Send the generated conjectures to an automated inductive theorem prover for proof, or output the result as a TIP-problem file. In the experiments reported in this paper, we use the aforementioned HipSpec system for proofs (\S \ref{sec:eval}).

\end{itemize}
This allows us to state the type class laws abstractly only once, and automatically infer and prove the concrete properties that any new instance need to satisfy to comply with the type class laws. 


\section{Background: HipSpec and TIP} 
\label{sec:background}
HipSpec allows the user to write down properties to prove in the Haskell source code, in a similar manner to how Haskell programmers routinely write QuickCheck properties. HipSpec supports a subset of the core Haskell language, with the caveat that functions are currently assumed to be terminating and values assumed to be finite. HipSpec can, if required, apply induction to the conjectures it is given, and then send the resulting proof obligations to an external theorem prover, such as Z3, E or Waldmeister \cite{z3,eprover,waldmeister}. The power of HipSpec comes from its \emph{theory exploration} phase: when given a conjecture to prove, HipSpec first use its subsystem QuickSpec \cite{quickspec,quickspec2}, which \emph{explores} the functions occurring in the problem by automatically suggesting a set of potentially useful basic lemmas, which HipSpec then proves by induction. These can then be used in the proof of the main conjecture. However, as theory exploration happens first, HipSpec sometimes also proves some extra properties, perhaps not strictly needed. We consider this a small price for the extra power theory exploration provides.

\example
\label{ex:oldHipSpec}
As a small example, consider asking HipSpec to prove that our \texttt{natAdd} function from the introduction indeed is commutative. 
The Haskell file is annotated with the property we wish to prove:

\begin{code}
prop_add_commute x y = x + y === y + x
\end{code}
The symbol \texttt{===} denote (polymorphic) equality in HipSpec properties. Calling HipSpec on this file instantly produces the output:

\begin{code}
Proved:
    (m + n) === (n + m)
    (m + (n + o)) === (n + (m + o))
    prop_add_commute  x + y === y + x 
\end{code}
Notice that HipSpec has printed out two properties it discovered and proved itself during its theory exploration phase (one of which is identical to the property we stated!). Proving \texttt{prop\_add\_commute} is then trivial. 



%
\subsubsection*{The TIP language.}
TIP is a general format for expressing problems for inductive theorem provers based on SMT-LIB \cite{smtlib}, extended with support for datatypes and recursive functions. The latest version of HipSpec also supports the TIP-language \cite{tip} as input in addition to Haskell. In our work, we use TIP as an intermediate language into which Haskell functions, types and properties are translated in order to facilitate passing them on to various external theorem provers.

We will not give an in depth description of the syntax of TIP here, save for a small example of the property we saw above, namely that addition is commutative:

\begin{code}
(assert-not (forall ((x Nat)) (= (plus x y) (plus y x))))
\end{code}
The keyword \texttt{assert-not} is used to tell the prover which properties to attempt to prove (this is inherited from SMT-LIB, where proofs are by refutation). Similarly, the keyword \texttt{assert} is used to tell the provers which properties are to be treated as axioms.

%
The ambition of TIP is to provide a shared format for many inductive theorem provers, thereby making it easier to share benchmarks and compare different provers. There exists a growing number of benchmarks for inductive theorem provers written in the TIP language, and a suite of tools for translating TIP into various common input formats for automated theorem provers such as SMT-LIB and TPTP \cite{tiptools,tptp}. We make use of the TIP language when instantiating type-class laws as described in the next section. The advantage is that this allows us to, in the future, more easily experiment with other inductive theorem proving backends, not only HipSpec. Furthermore, any inductive problems which the theorem prover cannot prove, can be added to the TIP benchmark suite to provide developers with additional challenge problems.

\section{Instantiating Type Class Laws}
In the previous section, we showed how HipSpec has traditionally been used, with the user annotating the source file with properties to prove. In a sense, HipSpec has been used mainly as a theorem prover, which just happened to take (a subset of) Haskell as input. 
In this work, we want to be able to handle more of the Haskell language, and thereby take a step towards building a more useful tool also for programmers, and not just as a theorem prover with Haskell as input.
%
%

%

\subsection{Expressing Type Class Laws}
\label{sec:expressing}
Type class law are expressed almost as normal HipSpec properties, using much of the same syntax as in Example \ref{ex:oldHipSpec}. The difference is that the type class laws use \emph{abstract functions}, i.e. functions which need to be supplied concrete definitions for the class instances. This is reflected in the type signature of the law, which contains type variables with class constraints. These abstract laws will give rise to multiple different versions specialised for each individual instance of that class. The type-class laws can be declared at the same time as the type-class itself, and later automatically instantiated when an instance of the corresponding type-class is declared.

\example 
\label{ex:tylaw}
Consider one of the type class laws for monoids which we have encountered before. As an abstract HipSpec property, it is defined as follows: 

\begin{code}
mappendAssoc :: Monoid a => a -> a -> a -> Equality a
mappendAssoc x y z = 
  mappend x (mappend y z) === mappend (mappend x y) z
\end{code}
Notice that the type variable \texttt{a} is required to be a \texttt{Monoid} in the type signature. \texttt{Equality} is HipSpec's type for equational properties to prove, and is represented using \texttt{===}. 

\subsection{Instantiating Laws}
\label{sec:spec}
HipSpec and TIP does currently only supports fully polymorphic type variables, i.e. type variables which have no constraints on the values they can assume. Since type class laws contain type variables with such constraints (e.g. the constraint that \texttt{a} must be a monoid in Example \ref{ex:tylaw}), they must be converted into specialised versions for each instance of the type class. This is rather straight-forward, and done by manipulating the GHC Core expressions resulting from compilation. For each type variable constrained by a type class, we first simply replace it with the type of all possible instances defined in the current file (if there are any). 

The type class law \texttt{mappendAssoc} from Example \ref{ex:tylaw} will for the \texttt{Nat} instance we saw before become automatically instantiated as shown below: 

\begin{code}
mappendAssoc1 :: Nat -> Nat -> Nat -> Equality Nat
mappendAssoc1 x y z = 
  mappend x (mappend y z) === mappend (mappend x y) z
\end{code} 
In the interest of readability of the example, we give the property in a Haskell-like syntax, rather than in GHC Core, but emphasise that this is part of an automated step in the implementation, and nothing the user will ever see as output. Notice that the type constraint \texttt{Monoid a} has disappeared.

GHC Core implements type classes by dictionary passing, i.e. when an overloaded function, like \texttt{mappend} is called, a lookup in the dictionary provides the appropriate concrete instance. For increased readability of the generated TIP-code, we inline dictionary lookups and rename them with more informative new names.

\begin{code}
MonoidNatmappend x (MonoidNatmappend y z) ===
    MonoidNatmappend (MonoidNatmappend x y) z
\end{code} 
The new function \texttt{MonoidNatmappend} is included in the resulting TIP file, defined as the corresponding function from Example \ref{ex:natmonoid} (i.e.  \texttt{+} or \texttt{*}, depending on which \texttt{Monoid} instance we are considering).

\subsection{Superclasses and Instances with Constraints}
\label{sec:axioms}
Sometimes, it is appropriate to include some additional information in the generated TIP problem file, other than the function definitions and instances of the type class laws. This includes, for example, assumptions that superclass laws are inherited by its subclasses and assumptions about laws that ought to hold about constrained type variables. Both cases are handled by introducing a new dummy type about which we can assert those extra facts.

\subsubsection{Superclass Laws.}
Haskell type classes may be defined as subclasses of already existing type classes. For instance, the type class \texttt{Ord}, used for totally ordered datatypes, is an extension of the type class for equality, \texttt{Eq}:

\begin{code}
class Eq a => Ord a where 
... 
\end{code}
This means that instances of \texttt{Ord} must also be instances of \texttt{Eq}, otherwise the compiler will complain. When generating instances of type class laws, we have therefore made the design decision to assume that all the superclass laws hold when attempting to prove the corresponding subclass laws. For example, this means assuming that the laws for \texttt{Eq} holds while proving the laws for a new instance of \texttt{Ord}. In other words, we assume that the \texttt{Eq} laws were already proved before, when \texttt{a} was declared an instance of \texttt{Eq}. This allows us to handle our proofs modularly, treating each type class in isolation and not having to worry about long chains of dependencies between type classes. 
The generated TIP file which we pass to the theorem prover must therefore include the super class laws as axioms holding for the type \texttt{a}. To achieve this, the type variable \texttt{a} is substituted by an arbitrary dummy datatype, about which we may include such assertions.

\subsubsection{Constrained Class Instances.}
Another similar scenario is when the class instance itself has a constrained type variable in its declaration. This is often the case for polymorphic data types, for example \texttt{Maybe a} if declaring it an instance of \texttt{Monoid} like in the example below:  

\begin{code}
instance Monoid a => Monoid (Maybe a) where
	...
\end{code}
In this case we need to take a little more care when instantiating the type class laws; we cannot simply replace the type variable \texttt{a} in the \texttt{Monoid a} constraint by all its possible instances, as this includes \texttt{Maybe a} itself, as well as \texttt{Maybe(Maybe(...))} and so on. 
Clearly this is not what's intended. Instead, we will interpret this to mean that: \emph{Assuming} that the type \texttt{a} is a \texttt{Monoid}, satisfying the associated type class laws, then prove that \texttt{Maybe a} does too. Just as for superclasses, we now substitute \texttt{a} for a new concrete dummy type, about which we can assert the laws for monoids, and use them when proving the laws for \texttt{Maybe}.



\section{Proving Type Class Laws}
\label{sec:eval}
Once the TIP files has been generated they are automatically piped through to an inductive theorem prover. Here we use HipSpec, but some other inductive theorem prover could equally well be integrated in our architecture as long as it supports TIP as input. We have focused the experiments presented here to a number of type classes where many of the laws are a little bit harder to prove than for basic type classes (such as \texttt{Eq}) and may require induction.  A longer evaluation on additional examples can be found in the MSc thesis accompanying this paper, but omitted here due to limited space \cite{msc}.


The experiments presented here use a development version of the HipSpec system, which includes improvements such as native support for the TIP-format as input, and options for more informative proof output which explains the steps of the proofs. However, unlike the previous version of HipSpec (described in \cite{hipspecCADE}), the new version does not, at the time of writing, fully support higher-order functions. We have therefore not yet been able to prove all laws of common type classes such as \texttt{Monoid}, \texttt{Functor} and \texttt{Monad}. Addressing this issue is ongoing work. 
Our tool can however output problem files in TIP-format also for type-classes containing higher-order functions.

\subsection{Experimental Results}
The timings in the experiments were obtained on a laptop computer with an Intel Core i7-3630QM 2.4 GHz processor and 8 GB of memory, running Arch Linux and GHC version 7.10.3. The source files for all the experiments are available online from: \url{https://github.com/chip2n/hipspec-typeclasses/src/Evaluation/}. 

In the experimental results, we use the symbol {\pass} to denote a successful proof, {\fail} to denote prover failure for a true law, and {\expectfail} to denote prover failure on a law violation, i.e. on a false instance of a law. HipSpec does not currently give counter-examples for laws that do not hold. For laws that cannot be proved, the user will have to use a separate tool like QuickCheck to check if a counter example can be found to indicate a law violation, as opposed to a true law which is beyond the provers automated capabilities. In the future, we would like to combine HipSpec with previous work on testing type class laws \cite{Jeuring2012}, to resolve this. 

\subsubsection{Semiring.}
Our first experiment comes from algebra, and is about semirings. A semiring is a set coupled with operators for addition (\texttt{+}) and multiplication (\texttt{*}), as well as identity values for both of them.
As a Haskell type class, this can be expressed as:

\begin{code}
class Semiring a where
    zero :: a
    one  :: a
    (+)  :: a -> a -> a
    (*)  :: a -> a -> a
\end{code}
The following eight laws are supposed to hold for semirings:
\begin{center}
  \begin{tabularx}{\textwidth}{| l  X | }
    \hline
      1. & \associativity{+}  \\
      2. & \identitytwo{+}{0} \\
      3. & \commutativity{+}  \\
      4. & \associativity{*}  \\
      5. & \identitytwo{*}{1} \\
      6. & \distributivityleft{*}{+}  \\
      7. & \distributivityright{*}{+} \\
      8. & \annihilation{*}{0} \\
    \hline
  \end{tabularx}
\end{center}
We experimented with the following four instances of semirings:
\begin{description}
  \item{\textbf{Nat}}: Natural numbers with addition and multiplication and identities 0 and 1.
  \item{\textbf{Bool (1)}}: Disjunction for addition and conjunction for multiplication with identities False and True
  \item{\textbf{Bool (2)}}: Conjunction for addition and disjunction for multiplication with identities True and False
  \item{\textbf{Matrix2}}: A 2-by-2 matrix with matrix addition and multiplication and identities empty matrix and identity matrix respectively, and entries belong to another semiring\footnote{The elements must belong to a semiring for the square matrix to do so.}.
\end{description}

\begin{center}
    \begin{tabular}{| l | c | c | c | c | c | c | c | c | c |}
      \multicolumn{10}{c}{\emph{Results: Semiring}} \\
      \hline
      \textbf{Instance} & \multicolumn{8}{c |}{\textbf{Law}} & \textbf{Total time} \\
      \hline
      & 1 & 2 & 3 & 4 & 5 & 6 & 7 & 8 & \\
      \hline
      Nat      & \pass & \pass & \pass & \pass & \pass & \pass & \pass & \pass & 28.2 s \\
      Bool (1) & \pass & \pass & \pass & \pass & \pass & \pass & \pass & \pass & 2.9 s \\
      Bool (2) & \pass & \pass & \pass & \pass & \pass & \pass & \pass & \pass & 2.9 s \\
      Matrix2  & \pass & \pass & \pass & \pass & \pass & \fail & \fail & \pass & 29.3 s \\
      \hline
    \end{tabular}
\end{center}
Most of these proofs are rather easy for HipSpec, except the distributivity properties for matrices (laws 7 and 8, which hold but are not proved here). We conjecture that this is due to the fact that HipSpec has not been told anything about auxiliary functions for summations of matrix elements row/column-wise, which is required here. Notice that in the case of the natural numbers and matrices, HipSpec spends more time doing theory exploration, and inventing lemmas. For the booleans, the proofs are straight-forward and do not even need induction.

\subsubsection{CommutativeSemiring.}
To demonstrate a type class which depend on a superclass, we also included a commutative semiring in our evaluation.

\begin{code}
class Semiring a => CommutativeSemiring a
\end{code}
As the name suggests, a commutative semiring is a semiring with one additional law stating that \texttt{*} is commutative:
\begin{center}
  \begin{tabularx}{\textwidth}{| l  X | }
    \hline
      1. & \commutativity{*}  \\
    \hline
  \end{tabularx}
\end{center}

We tested the same instances as for semirings:
\begin{center}
    \begin{tabular}{| l | c | c |}
      \multicolumn{3}{c}{\emph{Results: CommutativeSemiring}} \\
      \hline
      \textbf{Instance} & \textbf{Law} & \textbf{Total time} \\
      \hline
      & 1 & \\
      \hline
      Nat      & \pass & 21.0 s \\
      Bool (1) & \pass & 2.8 s \\
      Bool (2) & \pass & 2.5 s \\
      Matrix2  & \expectfail & 11.9 s \\
      \hline
    \end{tabular}
\end{center}
As expected, natural numbers and booleans can easily be shown to also be commutative semirings. Note that the matrix instance fails as expected; matrix multiplication is not commutative. 

\subsubsection{Reversible.}
The next type class characterise data structures that can be reversed: 

\begin{code}
class Reversible a where
    reverse :: a -> a
\end{code}
It has one law, stating that \texttt{reverse} is idempotent:
\begin{center}
  \begin{tabularx}{\textwidth}{| l  X | }
    \hline
      1. & $\mathit{reverse\ (reverse\ xs) = xs}$ \\
    \hline
  \end{tabularx}
\end{center}
We tested the following three instances:
\begin{description}
  \item{\textbf{List (rev)}}: Naive recursive list reverse.
  \item{\textbf{List (qrev)}}: List reverse using accumulator.
  \item{\textbf{Tree}}: A binary tree with a mirror operation, flipping the tree left to right. 
\end{description}

\begin{center}
    \begin{tabular}{| l | c | c |}
      \multicolumn{3}{c}{\emph{Results: Reversible}} \\
      \hline
      \textbf{Instance} & \textbf{Law} & \textbf{Total time} \\
      \hline
       & 1 & \\
       \hline
      List (\emph{rev})  & \pass  & 5.9 s \\
      List (\emph{qrev}) & \pass  & 5.7 s \\
      Tree               & \pass &  0.9 s \\
      \hline
    \end{tabular}
\end{center}
All these proofs require induction, and the ones about lists also need auxiliary lemmas which HipSpec must discover and prove first. However, as there is only very few functions and laws present (unlike for semirings), this is rather fast.

\subsubsection{Monoid.}
We have already encountered the Monoid type class as a running example. 
 \begin{center}
  \begin{tabularx}{\textwidth}{| l  X | }
    \hline
         1. & $\mathit{mappend\ mempty\ x = x}$\\
         2. & $\mathit{mappend\ x\ mempty = x}$\\
         3. & $\mathit{mappend\ x\ (mappend\ y\ z) = mappend\ (mappend\ x\ y)\ z}$\\
    \hline
  \end{tabularx}
\end{center}
We here omit the 4th law (the default implementation for \texttt{mconcat}) which is higher-order and, as mentioned above, not yet supported in the development version of HipSpec we are using. We give the results for the remaining three laws, and do so for the datatype instances we have seen before. The \texttt{Matrix2} instance has the additional constraint of requiring its elements to also be monoids, and similarly, \texttt{Maybe a} has a constraint requiring \texttt{a} to be a monoid. 

\begin{center}
    \begin{tabular}{| l | c | c | c | c | c |}
      \multicolumn{5}{c}{\emph{Results: Monoid}} \\
      \hline
      \textbf{Instance} & \multicolumn{3}{c |}{\textbf{Law}} & \textbf{Total time} \\
      \hline
      & 1 & 2 & 3 & \\
      \hline
      Nat (\emph{add}) & \pass & \pass & \pass & 4.3 s \\
      Nat (\emph{mul}) & \pass & \pass & \pass & 27.3 s \\
      Matrix2 (\emph{mul})  & \pass & \pass & \pass & 2.3 s \\
      List             & \pass & \pass & \pass & 1.5 s \\
      Maybe            & \pass & \pass & \pass & 2.1 s \\
      \hline
    \end{tabular}
\end{center}
Laws 1 and 2 for monoids are rather trivial. The only instance needing slightly longer is when $\mathit{mappend}$ is instantiated to natural number  multiplication, in which law 3 needs some lemmas to be discovered and proved.

\subsection{Discussion of Results and Further Work}
The proof obligations resulting from the type classes above are mainly within the comfort zone of an inductive theorem prover such as HipSpec. Usually, the prover just needs a few seconds to complete the proofs, unless it has discovered many candidate lemmas during the theory exploration phase. This is the downside of using HipSpec: it sometimes eagerly spends some time proving lemmas that turn out not to be necessary for the final proof. Targeting the exploration by for example attempting to extract information from failed proofs is further work.

It is somewhat disappointing that we have not yet been able to automatically prove any higher-order laws, but hope that it is merely a matter of time until support has been added also to the new version of HipSpec. The lack of higher-order support is not inherent to the new version of HipSpec, but due to an incomplete module in HipSpec for reading in TIP files. We could have opted for using an old version of HipSpec, but it does not support TIP as an input language, and we would have had to instead produce modified copies of the original Haskell source files to give to the prover. This would likely have meant less automation of the whole proving pipeline. Furthermore, the new version of HipSpec uses the latest version of QuickSpec for lemma discovery \cite{quickspec2}, which is much faster and more powerful that the old version.
We also think it is valuable to output the instantiated laws and proof obligations in the TIP format, which can more easily be shared between different theorem proving systems. This means that we are not actually bound to just a single theorem prover (HipSpec), but can quite easily connect our work also to other provers, should a better one become available. Furthermore, the user could edit the TIP file output by our tool to for example add extra facts that the prover did not discovery automatically. 

HipSpec's proof output depends on which external prover it uses for proof obligations. Details of the rewriting steps can be presented to the user when Waldmeister is used \cite{waldmeister}, but we often want to use other more powerful provers such as Z3 \cite{z3}. Adding richer proof output for additional prover backends, counter-examples for laws violations and other user feedback about failed proofs is further work. This has however been partially addressed in HipSpec's sister system Hipster, which shares the same theory exploration component, but is integrated in the interactive proof assistant Isabelle/HOL \cite{isabelle,hipster} for proofs. Hipster produces snippets of Isabelle proof scripts for discovered lemmas. 

HipSpec currently assumes that functions to be terminating and values are finite. There is however nothing in principle that would stop us from, for example, connecting HipSpec to a termination checker and extending it with for instance co-induction to reason about infinite structures. 

\section{Related Work}
Zeno is an earlier inductive theorem prover for Haskell \cite{zeno}, which is unfortunately no longer maintained. Unlike HipSpec it had its own internal theorem prover, while HipSpec is designed around the TIP language for easier communication with different external provers used to handle proof obligations arising after applying induction. Zeno did not support theory exploration but instead conjectured lemmas based subgoals remaining in stuck proofs.

HERMIT \cite{Farmer2015} is a system for mechanising equational reasoning about Haskell programs.
Properties are written as GHC rewrite rules in the Haskell source code and proven either interactively in the HERMIT shell or by writing a script which automates the process. As opposed to HipSpec, which is fully automatic and exploits external theorem provers, HERMIT is an more of an interactive proof assistant relying on the user to specify the steps of proofs. HERMIT also requires the user to supply all lemmas, and does not support any form of theory exploration like HipSpec.

LiquidHaskell is a contract-based verification framework based on refinement types \cite{Vazou2014}. Refinement types are essentially the addition of logical predicates to `refine' the types of inputs and outputs to functions. These predicates are restricted to theories with decision procedures, which ensures contracts are written in a logic suitable for being checked by a SMT solver. LiquidHaskell is a more mature system than HipSpec, but restricted to decidable theories so it does not support automated induction and theory exploration.   

\section{Conclusion}
This work demonstrates how automated theorem proving and functional programming can be integrated to provide benefit to programmers. We showed how type class laws can be expressed as abstract properties, then instantiated and proved when we declare a new instance of a type class. Moving type class laws from informal descriptions in the documentation to something that is actually enforced would both help programmers to a shared understanding of the purpose of the type class as well as helping them implement instances correctly. 

Many Haskell programmers are today routinely writing QuickCheck properties which are automatically tested on many randomly generated values. In the future, we envisage programmers writing similar properties and laws, and using a tool similar to what's described here to not only test, but also \emph{prove} their programs correct, with little more overhead than using QuickCheck. We believe our work is at least a small step on the way towards this ambitious goal.

\bibliographystyle{plain}
\bibliography{paper}

\end{document}